\documentstyle[12pt]{article}

\catcode`\@=11
\long\def\@makefntext#1{ 
\protect\noindent \hbox to 3.2pt {\hskip-.9pt
$^{{\ninerm\@thefnmark}}$\hfil}#1\hfill} 

\def\thefootnote{\fnsymbol{footnote}}
 \def\@makefnmark{\hbox to 0pt{$^{\@thefnmark}$\hss}}  

\def\ps@myheadings{\let\@mkboth\@gobbletwo
\def\@oddhead{\hbox{} 
\rightmark\hfil\ninerm\thepage}
\def\@oddfoot{}\def\@evenhead{\ninerm\thepage\hfil 
\leftmark\hbox{}}\def\@evenfoot{}
\def\sectionmark##1{}\def\subsectionmark##1{}}

\begin{document}

\newcommand{\symbolfootnote}{\renewcommand{\thefootnote}
        {\fnsymbol{footnote}}}
\renewcommand{\thefootnote}{\fnsymbol{footnote}}
\newcommand{\alphfootnote}
        {\setcounter{footnote}{0}
         \renewcommand{\thefootnote}{\sevenrm\alph{footnote}}}

\newcounter{sectionc}\newcounter{subsectionc}\newcounter{subsubsectionc}
\renewcommand{\section}[1] {\vspace{0.6cm}\addtocounter{sectionc}{1}
\setcounter{subsectionc}{0}\setcounter{subsubsectionc}{0}\noindent
        {\bf\thesectionc. #1}\par\vspace{0.4cm}}
\renewcommand{\subsection}[1] {\vspace{0.6cm}\addtocounter{subsectionc}{1}
        \setcounter{subsubsectionc}{0}\noindent
        {\it\thesectionc.\thesubsectionc. #1}\par\vspace{0.4cm}}
\renewcommand{\subsubsection}[1]
{\vspace{0.6cm}\addtocounter{subsubsectionc}{1}
        \noindent {\rm\thesectionc.\thesubsectionc.\thesubsubsectionc.
        #1}\par\vspace{0.4cm}}
\newcommand{\nonumsection}[1] {\vspace{0.6cm}\noindent{\bf #1}
        \par\vspace{0.4cm}}

\newcounter{appendixc}
\newcounter{subappendixc}[appendixc]
\newcounter{subsubappendixc}[subappendixc]
\renewcommand{\thesubappendixc}{\Alph{appendixc}.\arabic{subappendixc}}
\renewcommand{\thesubsubappendixc}
        {\Alph{appendixc}.\arabic{subappendixc}.\arabic{subsubappendixc}}

\renewcommand{\appendix}[1] {\vspace{0.6cm}
        \refstepcounter{appendixc}
        \setcounter{figure}{0}
        \setcounter{table}{0}
        \setcounter{equation}{0}
        \renewcommand{\thefigure}{\Alph{appendixc}.\arabic{figure}}
        \renewcommand{\thetable}{\Alph{appendixc}.\arabic{table}}
        \renewcommand{\theappendixc}{\Alph{appendixc}}
        \renewcommand{\theequation}{\Alph{appendixc}.\arabic{equation}}
        \noindent{\bf Appendix \theappendixc #1}\par\vspace{0.4cm}}
\newcommand{\subappendix}[1] {\vspace{0.6cm}
        \refstepcounter{subappendixc}
        \noindent{\bf Appendix \thesubappendixc. #1}\par\vspace{0.4cm}}
\newcommand{\subsubappendix}[1] {\vspace{0.6cm}
        \refstepcounter{subsubappendixc}
        \noindent{\it Appendix \thesubsubappendixc. #1}
        \par\vspace{0.4cm}}

\def\abstracts#1{{
        \centering{\begin{minipage}{30pc}\tenrm\baselineskip=12pt\noindent
        \centerline{\tenrm ABSTRACT}\vspace{0.3cm}
        \parindent=0pt #1
        \end{minipage} }\par}}

\newcommand{\bibit}{\it}
\newcommand{\bibbf}{\bf}
\renewenvironment{thebibliography}[1]
        {\begin{list}{\arabic{enumi}.}
        {\usecounter{enumi}\setlength{\parsep}{0pt}
\setlength{\leftmargin 1.25cm}{\rightmargin 0pt}
         \setlength{\itemsep}{0pt} \settowidth
        {\labelwidth}{#1.}\sloppy}}{\end{list}}

\topsep=0in\parsep=0in\itemsep=0in
\parindent=1.5pc

\newcounter{itemlistc}
\newcounter{romanlistc}
\newcounter{alphlistc}
\newcounter{arabiclistc}
\newenvironment{itemlist}
        {\setcounter{itemlistc}{0}
         \begin{list}{$\bullet$}
        {\usecounter{itemlistc}
         \setlength{\parsep}{0pt}
         \setlength{\itemsep}{0pt}}}{\end{list}}

\newenvironment{romanlist}
        {\setcounter{romanlistc}{0}
         \begin{list}{$($\roman{romanlistc}$)$}
        {\usecounter{romanlistc}
         \setlength{\parsep}{0pt}
         \setlength{\itemsep}{0pt}}}{\end{list}}

\newenvironment{alphlist}
        {\setcounter{alphlistc}{0}
         \begin{list}{$($\alph{alphlistc}$)$}
        {\usecounter{alphlistc}
         \setlength{\parsep}{0pt}
         \setlength{\itemsep}{0pt}}}{\end{list}}

\newenvironment{arabiclist}
        {\setcounter{arabiclistc}{0}
         \begin{list}{\arabic{arabiclistc}}
        {\usecounter{arabiclistc}
         \setlength{\parsep}{0pt}
         \setlength{\itemsep}{0pt}}}{\end{list}}

\newcommand{\fcaption}[1]{
        \refstepcounter{figure}
        \setbox\@tempboxa = \hbox{\tenrm Fig.~\thefigure. #1}
        \ifdim \wd\@tempboxa > 6in
           {\begin{center}
        \parbox{6in}{\tenrm\baselineskip=12pt Fig.~\thefigure. #1 }
            \end{center}}
        \else
             {\begin{center}
             {\tenrm Fig.~\thefigure. #1}
              \end{center}}
        \fi}

\newcommand{\tcaption}[1]{
        \refstepcounter{table}
        \setbox\@tempboxa = \hbox{\tenrm Table~\thetable. #1}
        \ifdim \wd\@tempboxa > 6in
           {\begin{center}
        \parbox{6in}{\tenrm\baselineskip=12pt Table~\thetable. #1 }
            \end{center}}
        \else
             {\begin{center}
             {\tenrm Table~\thetable. #1}
              \end{center}}
        \fi}

\def\@citex[#1]#2{\if@filesw\immediate\write\@auxout
        {\string\citation{#2}}\fi
\def\@citea{}\@cite{\@for\@citeb:=#2\do
        {\@citea\def\@citea{,}\@ifundefined
        {b@\@citeb}{{\bf ?}\@warning
        {Citation `\@citeb' on page \thepage \space undefined}}
        {\csname b@\@citeb\endcsname}}}{#1}}

\newif\if@cghi
\def\cite{\@cghitrue\@ifnextchar [{\@tempswatrue
        \@citex}{\@tempswafalse\@citex[]}}
\def\citelow{\@cghifalse\@ifnextchar [{\@tempswatrue
        \@citex}{\@tempswafalse\@citex[]}}
\def\@cite#1#2{{$\null^{#1}$\if@tempswa\typeout
        {IJCGA warning: optional citation argument
        ignored: `#2'} \fi}}
\newcommand{\citeup}{\cite}

\def\fnm#1{$^{\mbox{\scriptsize #1}}$}
\def\fnt#1#2{\footnotetext{\kern-.3em
        {$^{\mbox{\sevenrm #1}}$}{#2}}}

\font\twelvebf=cmbx10 scaled\magstep 1
\font\twelverm=cmr10 scaled\magstep 1
\font\twelveit=cmti10 scaled\magstep 1
\font\elevenbfit=cmbxti10 scaled\magstephalf
\font\elevenbf=cmbx10 scaled\magstephalf
\font\elevenrm=cmr10 scaled\magstephalf
\font\elevenit=cmti10 scaled\magstephalf
\font\bfit=cmbxti10
\font\tenbf=cmbx10
\font\tenrm=cmr10
\font\tenit=cmti10
\font\ninebf=cmbx9
\font\ninerm=cmr9
\font\nineit=cmti9
\font\eightbf=cmbx8
\font\eightrm=cmr8
\font\eightit=cmti8


\centerline{\tenbf OUR PRESENT UNDERSTANDING OF CP VIOLATION}
\vspace{0.8cm}
\centerline{\tenrm JONATHAN L. ROSNER}
\baselineskip=13pt
\centerline{\tenit Enrico Fermi Institute and Department of Physics
University of Chicago}
\baselineskip=12pt
\centerline{\tenit 5640 S. Ellis Ave.~, Chicago, IL 60637, USA}
\vspace{-1.7in}
\rightline{EFI 94-25}
\rightline{hep-ph/9407257}
\rightline{July 1994}
\vspace{-0.5in}
\leftline{Presented at PASCOS 94 Conference}
\leftline{Syracuse, NY, May 19 -- 24, 1994}
\leftline{Proceedings to be published by World Scientific}
\vspace{1.2in}

\vspace{0.9cm}
\abstracts{If CP violation in the decays of neutral kaons is due to phases in
the weak couplings of quarks, as encoded in the Cabibbo-Kobayashi-Maskawa (CKM)
matrix, there are many other experimental consequences.  Notable among these
are CP-violating rate asymmetries and triangle relations among decay rates in
$B$ meson decays, while charmed particle decays should not be a good place to
see CP-violating effects.  In the context of the CKM and other models of CP
violation, we discuss phenomena such as electric dipole moments, the baryon
asymmetry of the Universe, and the strong CP problem, and speculate on a common
origin for CP-violating phenomena.}

\vfil
\twelverm
\baselineskip=14pt
\section{Introduction}

A crucial step in the formulation of the theory of weak interactions was the
realization that parity (P) and charge conjugation (C) were not
conserved.\cite{LY}  The theory as formulated\cite{VA} in 1957 did conserve the
product CP.  Seven years later, the discovery of the two-pion decay of the
neutral kaon\cite{CCFT} showed that even the product CP was violated.  Since
1964, although no new CP-violating phenomena have been observed, we have a
theory for this effect and the prospect of many experimental tests. This review
describes our present understanding of CP violation and some of the tests which
are likely to bear fruit in the near future.

We begin in Section 2 by discussing present information on CP violation in the
system of neutral kaons.  Although no direct evidence bearing on the CKM theory
has been obtained, there are new results exhibiting greater consistency with
invariance under the product CPT, where T stands for time reversal.  This is
encouraging, since once cannot write a CPT-violating theory without violating
cherished properties of quantum field theory.\cite{LP}

The leading candidate for CP violation in the kaon system is the presence of
phases in the CKM matrix, described in Section 3. Present information allows
one to specify a range of parameters for which the theory provides a
self-consistent description of both $B - \bar B$ and CP-violating $K - \bar K$
mixing.  Ways in which these parameters can be specified more precisely are
described in Section 4. These include further studies of kaons, and numerous
experiments on $B$ mesons treated separately in Section 5.

Even though the CKM theory works so well, there are alternative possibilities
for CP violation, some of which are described in Section 6. Although the
standard CKM picture predicts minuscule effects of CP violation in the charm
sector, as mentioned in Section 7, these effects could be larger in some of the
other theories (notably multi-Higgs models).

Although the only direct manifestation of CP violation appears in the neutral
kaon system, there is indirect evidence from the existence of a non-zero
baryon number in the Universe.\cite{Sakh}  In Section 8 we comment on one
likely scenario for the origin both of the CP violation in the kaon system and
the the CP violation needed to produce an excess of baryons over antibaryons.

One place in which CP violation could have shown up, but doesn't, is in the
strong interactions.  The structure of the vacuum could have led to measurable
CP-violating effects, for example in the electric dipole moment of the neutron.
One way to avoid such effects is through a spontaneously broken
symmetry\cite{PQ} and the existence of a light pseudoscalar particle.\cite{AX}
Section 9 contains some brief remarks on searches for this particle.

We summarize in Section 10.

\section{The kaon system}\label{sec:kaons}

\subsection{States of definite lifetime}

The strangeness hypothesis requires there to be two distinct neutral
kaons:  a $K^0$ and a $\bar K^0$.  Since they are degenerate, how can one tell
them apart?  The weak interactions split linear combinations of these states
into ones with definite mass and lifetime. In the limit of CP conservation, one
state, $K_1 \equiv (K^0 + \bar K^0)/\sqrt{2}$, with even CP, can couple to the
CP-even $2 \pi$ state, and thus is short-lived. The other, $K_2 \equiv (K^0 -
\bar K^0)/\sqrt{2}$, with odd CP, is long-lived, since it cannot decay to $2
\pi$ but only to three-body final states.\cite{GP}

The discovery\cite{CCFT} in 1964 that the long-lived neutral kaon also decays
to
$2 \pi$ showed that CP is not a valid symmetry of the weak interactions.  The
short-lived and long-lived states of definite mass and lifetime can be written
as
\begin{equation}
K_S \simeq K_1 + \epsilon K_2~~,~~~
K_L \simeq K_2 + \epsilon K_1~~,~~~
\end{equation}
where $\epsilon$ encodes all we know at present about CP violation. It is
helpful to discuss briefly the phenomenology of the neutral kaon system.  Much
more complete discussions may be found, for example, in several
reviews.\cite{revs}

\subsection{Kaon mixing and phases}

In a two-component basis labeled by $K^0$ and $\bar K^0$, the mass matrix
$\cal{M}$ has eigenstates $|S \rangle \equiv |K_S \rangle$ and $|L \rangle
\equiv |K_L \rangle$ with corresponding complex eigenvalues $\mu_{S,L} =
m_{S,L}  - i \gamma_{S,L}/2$. One can decompose $\cal{M}$ in terms of two
Hermitian matrices $M$ and $\Gamma$:  ${\cal M} = M - i \Gamma/2$.  The
requirement of CPT invariance (which we shall assume here) implies ${\cal
M}_{11} = {\cal M}_{22}$.  A detailed investigation of final states in
neutral kaon decays\cite{BS} shows that one can write
$\epsilon \simeq i~{\rm Im}~M_{12}/(\mu_S - \mu_L)$;
an additional contribution from Im $\Gamma_{12}$ turns out to be neglible.
Consequently, the measured differences between masses and lifetimes of the
weak eigenstates allow one to conclude that Arg $\epsilon = 43.3^{\circ}$.
Studies of CP-violating kaon decays have shown that $|\epsilon| = (2.26
\pm 0.02) \times 10^{-3}$.

The ratios of amplitudes for long-lived and short-lived decays of kaons to
two pions are
\begin{equation}
\eta_{+-} = \frac{A(K_L \to \pi^+ \pi^-)}{A(K_S \to \pi^+ \pi^-)}
= \epsilon + \epsilon'~~,~~~
\eta_{00} = \frac{A(K_L \to \pi^0 \pi^0)}{A(K_S \to \pi^0 \pi^0)}
= \epsilon - 2 \epsilon'~~~,
\end{equation}
where
\begin{equation}
\epsilon' = \frac{i~{\rm Im}~A_s}{\sqrt{2} A_0}e^{i(\delta_2 - \delta_0)}~~~.
\end{equation}

Here the subscripts $I = 0,~2$ on the amplitudes $A_I$ and the elastic phase
shifts $\delta_I$ refer to the isospin of the $\pi \pi$ system.  The measured
phase shifts\cite{phs} imply that Arg $\epsilon' = (43 \pm 6)^{\circ}$,
very close to the expected phase of $\epsilon$.  Consequently, one expects
nearly maximal constructive or destructive interference between $\epsilon$
and $\epsilon'$ in the ratios of decay rates for $K_L$ and $K_S$ to decay
to pairs of charged and neutral pions.

The quantity Arg $\epsilon = 43.3^{\circ}$ is sometimes called the
``superweak'' phase, since it is that which $\eta_{+-}$ and $\eta_{00}$ would
have in a theory\cite{sw} in which CP violation arose purely from a
``superweak'' mixing between $K^0$ and $\bar K^0$.  Since the phase of
$\epsilon'$ is expected to be so close to that of $\epsilon$, and since
the magnitude $\epsilon'/\epsilon$ is smaller in any case than a few parts
in $10^3$ (as will be discussed in Sec.~4.5), deviations of the phases
$\Phi_{+-} \equiv {\rm Arg}~\eta_{+-}$ and $\Phi_{00} \equiv {\rm
Arg}~\eta_{00}$ from the superweak value are expected to be extremely small if
CPT invariance is valid. (Parametrizations in which CPT violation is allowed
have been discussed, for example, in several reviews.\cite{CPTv})

\subsection{Recent developments}

A recent experiment at Fermilab\cite{E773} has measured the phase $\Phi_{+-}$
to be very close to the ``superweak'' value:  $\Phi_{+-} = (43.35 \pm 0.70
\pm 0.79)^{\circ}$.  The difference $\Delta \Phi \equiv \Phi_{+-} - \Phi_{00}$
is measured even more precisely, since the phase of the $K_S$ regeneration
amplitude cancels: $\Delta \Phi = (0.67 \pm 0.85 \pm 1.1)^{\circ}$.  This is
in marked contrast to the situation several years ago, when $\Delta \Phi$
appeared to be about two standard deviations away from zero with an error
of about six degrees.

Another indication of the consistency of $\Phi_{+-}$ with expectations is the
quantity 2 Re $\epsilon = (3.30 \pm 0.12) \times 10^{-3}$ measured via the
charge asymmetry in $K_L \to \pi \ell \nu$ decays.\cite{kl3}  Using the new
value of $\Phi_{+-}$, and neglecting the small contribution of $\epsilon'$, we
obtain from the measured values of $|\epsilon|$ and $\Phi_{+-}$ the value 2 Re
$\epsilon = (3.29 \pm 0.07) \times 10^{-3}$, in excellent agreement with that
measured directly.

These results, together with the inconclusive nature of the search for
$\epsilon' \ne 0$ to be mentioned in Sec.~4.5, leave the parameter $\epsilon$
as the single manifestation of CP violation for which we have hard evidence
so far.  CPT appears for the moment to be a valid symmetry.  We now discuss a
possible reason for $\epsilon \ne 0$.

\section{CKM Phases}\label{sec:CKM}

\subsection{Candidate theory of CP violation}

The box diagrams which mix $K^0 \equiv d \bar s$ and $\bar K^0 \equiv s \bar d$
involve intermediate states of $u,~c,~t$ quarks and the corresponding
charge-changing couplings of $W$ bosons.  Phases in these couplings can give
rise to a CP-violating mixing term, leading to $\epsilon \ne 0$.  Three quark
families are needed in order to obtain phases which cannot be rotated away by
redefinition of quark fields.\cite{KM}

Neutral $B$ mesons also mix with their antiparticles, and since they involve
bottom rather than strange quarks, some of the phases in the couplings are
different.

\subsection{Origin of the CKM matrix}

The initial SU(2)$_L \times$ U(1) electroweak Lagrangian, before electroweak
symmetry breaking, may be written in flavor-diagonal form as
\begin{equation}
{\cal L}_{\rm int} = - \frac{g}{\sqrt{2} }[ \overline{U '}_L
\gamma^\mu W_\mu^{(+)} {D'}_L + H.c.]~~~~~,
\end{equation}
where $U' \equiv (u',c',t')$ and $D' \equiv (d',s',b')$ are column vectors
decribing {\em weak eigenstates}. Here $g$ is the weak $SU(2)_L$ coupling
constant, and $\psi_L \equiv (1 - \gamma_5 ) \psi /2$ is the left-handed
projection of the fermion field $\psi = U$ or $D$.

Quark mixings arise because mass terms in the Lagrangian are permitted to
connect weak eigenstates with one another. Thus, the matrices ${\cal M}_{U,~D}$
in
\begin{equation}
{\cal L}_m = - [\overline{U '}_R {\cal M}_U {U'}_L + \overline {D '}_R {\cal
M}_D {D'}_L + H.c.]
\end{equation}
may contain off-diagonal terms. One may diagonalize these matrices by separate
unitary transformations on left-handed and right-handed quark fields:
\begin{equation}
R_{Q}^+ {\cal M}_{Q} L_{Q} = L_{Q}^+ {\cal M}_{Q}^+ R_Q = \Lambda_Q ~~~~~.
\end{equation}
where
\begin{equation}
{Q'}_L = L_Q Q_L ; ~~ {Q'}_R = R_Q Q_R ~~~ (Q = U, D)~~~~~~.
\end{equation}
Using the relation between weak eigenstates and mass eigenstates:
${U'}_L = L_U U_L , ~ {D'}_L = L_D D_L$, we find
\begin{equation}
{\cal L}_{\rm int} = - \frac{g}{\sqrt{2}} [ \overline{U}_L \not{W} V D_L +
H.c.] ~~~~,
\end{equation}
where $U \equiv (u,c,t)$ and $D \equiv (d,s,b)$ are the mass eigenstates, and
$V \equiv L_U^+ L_D$. The matrix $V$ is just the Cabibbo-Kobayashi-Maskawa
matrix. By construction, it is unitary: $V^+V = VV^+ = 1$. It carries no
information about $R_U$ or $R_D$. More information would be forthcoming from
interactions sensitive to right\--handed quarks or from a genuine theory of
quark masses.

\subsection{Parameters and their values}

The CKM matrix for three families of quarks and leptons will have four
independent parameters no matter how it is represented. In a
parametrization\cite{wp} in which the rows of the CKM matrix are labelled by
$u,~c,~t$ and the columns by $d,~s,~b$, we may write
\begin{equation}
V = \left ( \begin{array}{c c c}
V_{ud} & V_{us} & V_{ub} \\
V_{cd} & V_{cs} & V_{cb} \\
V_{td} & V_{ts} & V_{tb}
\end{array} \right )
\approx \left [ \matrix{1 - \lambda^2 /2 & \lambda & A \lambda^3 ( \rho -
i \eta ) \cr
- \lambda & 1 - \lambda^2 /2 & A \lambda^2 \cr
A \lambda^3 ( 1 - \rho - i \eta ) & - A \lambda^2 & 1 \cr } \right ]~~~~~ .
\end{equation}
Note the phases in the elements $V_{ub}$ and $V_{td}$.  These phases allow the
standard $V - A$ interaction to generate CP violation as a higher-order weak
effect.

The parameters of the CKM matrix are measured in various ways, some of which
are described in more detail elsewhere at this conference.\cite{stonepas}

\begin{enumerate}

\item The parameter $\lambda$ is measured by a comparison of strange particle
decays with muon decay and nuclear beta decay, leading to $\lambda \approx
\sin \theta \approx 0.22$, where $\theta$ is just the Cabibbo\cite{cab} angle.

\item The dominant decays of $b$-flavored hadrons occur via the element
$V_{cb} = A \lambda^2$.  The lifetimes of these hadrons and their semileptonic
branching ratios then lead to estimates in the range $A = 0.7 - 0.9$.

\item The decays of $b$-flavored hadrons to charmless final states allow one
to measure the magnitude of the element $V_{ub}$ and thus to conclude that
$\sqrt{\rho^2 + \eta^2} = 0.2 - 0.5$.

\item The least certain quantity is the phase of $V_{ub}$:  Arg $(V_{ub}^*)
= \arctan(\eta/\rho)$.  We shall mention ways in which information on this
quantity may be improved, in part by indirect information associated with
contributions of higher-order diagrams involving the top quark.

\end{enumerate}

The unitarity of V and the fact that $V_{ud}$ and $V_{tb}$ are very close to
1 allows us to write $V_{ub}^* + V_{td} \simeq A \lambda^3$, or, dividing
by a common factor of $A \lambda^3$,
\begin{equation}
\rho + i \eta ~~ + ~~ (1 - \rho - i \eta) = 1~~~.
\end{equation}
The point $(\rho,\eta)$ thus describes in the complex plane one vertex of a
triangle\cite{UT} whose other two vertices are $(0,0)$ and $(0,1)$.  This
triangle and conventional definitions of its angles are depicted in Fig.~1.

\begin{figure}
\vspace{1.5in}
\caption{The unitarity triangle.  (a) Relation obeyed by CKM elements; (b)
relation obeyed by (CKM elements)/$A \lambda^3$}
\end{figure}

\subsection{Indirect information}

Indirect information on the CKM matrix comes from $B^0 - \bar B^0$ mixing and
CP-violating $K^0 - \bar K^0$ mixing, through the contributions of box diagrams
involving two charged $W$ bosons and two quarks of charge 2/3 $(u,~c,~t)$ on
the intermediate lines.  Evidence for the top quark with a mass of $m_t = 174
\pm 10~^{+13}_{-12}$ GeV/$c^2$ has recently been reported,\cite{CDFtop}
reducing the errors associated with these box diagrams.

The original evidence for $B^0 - \bar B^0$ mixing came from the presence of
``wrong-sign'' leptons in $B$ meson semileptonic decays.\cite{bbmix}  The
splitting $\Delta m_B$ between mass eigenstates is proportional to $f_B^2 m_t^2
|V_{td}|^2$ times a slowly varying function of $m_t$.  Here $f_B$ is the $B$
meson decay constant.  The contributions of lighter quarks in the box diagrams,
while necessary to cut off the high-energy behavior of the loop integrals, are
numerically insignificant.

The CKM element $|V_{td}|$ is proportional to $|1 - \rho - i \eta|$.  Thus,
exact knowledge of $\Delta m_B,~f_B$ and $m_t$ would specify a circular arc
in the $(\rho,\eta)$ plane with center (1,0).  Errors on all these quantities
spread this arc out into a band, as illustrated schematically by the dashed
lines in Fig.~2.

\begin{figure}
\vspace{1.75in}
\caption{Types of allowed regions in $(\rho,\eta)$ plane arising from $B -
\bar B$ mixing (dashes), $CP$-violating $K - \bar K$ mixing (solid), and
$|V_{ub}/V_{cb}|$ (dots)}
\end{figure}

Present averages\cite{mixavg} give $(\Delta m_B/\Gamma_B) = 0.71 \pm 0.07$.
This large value is good news for the prospect of observing CP-violating
asymmetries in $B^0$ decays.

Similar box diagrams govern CP-violating $K^0 - \bar K^0$ mixing.  Here the
dominant contribution to the imaginary part of the mass matrix element $M_{12}$
discussed in Sec.~2, which gives rise to the parameter $\epsilon$, is
proportional to $f_K^2 m_t^2$ Im $(V_{td}^2)$ times a slowly varying function
of $m_t$.  Charmed quarks also provide a small contribution.

The kaon decay constant is known: $f_K = 160$ MeV.  The imaginary part of
$V_{td}$ is proportional to $\eta(1-\rho)$.  Knowledge of $\epsilon$ thus
specifies a hyperbola in the $(\rho,\eta)$ plane with focus at $(1,0)$, which
is spread out into a band (the solid lines in Fig.~2) because of uncertainties
in hadronic matrix elements.

\subsection{Allowed $(\rho,\eta)$ region}

\begin{figure}
\vspace{3.4in}
\caption{Contours of 68\% (inner curve) and 90\% (outer curve) confidence
levels for regions in the $(\rho,\eta)$ plane.  Dotted semicircles denote
central value and $\pm 1 \sigma$ limits implied by $|V_{ub}/V_{cb}| = 0.08 \pm
0.03$.  Plotted point corresponds to minimum $\chi^2 = 0.17$, while (dashed,
solid) curves correspond to $\Delta \chi^2 = (2.3,~4.6)$}
\end{figure}

Information on $|V_{ub}/V_{cb}|$ specifies a circular band in the $(\rho,\eta)$
plane, as depicted by the dotted lines in Fig.~2.  When this constraint is
added to those mentioned in the previous subsection, one obtains the
potato-shaped region shown in Fig.~3.  Here we have taken $m_t = 174 \pm 17$
GeV/$c^2$, $f_B = 180 \pm 30$ MeV, $(\rho^2 + \eta^2 )^{1/2} = 0.36 \pm 0.14$
(corresponding to $|V_{ub}/V_{cb}| = 0.08 \pm 0.03$), and $A = 0.79 \pm 0.09$
(corresponding to $V_{cb} = 0.038 \pm 0.005$). A parameter known as $B_K$
describes the degree to which the box diagrams dominate the $CP$-violating $K^0
- \bar K^0$ mixing.  We take $B_K = 0.8 \pm 0.2$, and
set the corresponding value for $B$ mesons equal to 1. A QCD correction to the
$B^0 - \bar B^0$ mixing amplitude has been taken to be $\eta_{\rm QCD} = 0.6
\pm 0.1$; this is perhaps a generous error in view of the existence of a more
precise estimate of this quantity.\cite{BB}  Other parameters and fitting
methods are as discussed in more extensive treatments elsewhere.\cite{HR,CKM}
Several parallel analyses\cite{BUR,AL} reach qualitatively similar conclusions.

The best fit corresponds to $\rho \simeq 0,~\eta \simeq 0.36$, while at
90\% confidence level the allowed ranges are:
$$
\eta \simeq 0.3~~:~~~-0.4 \le \rho \le 0.4~~~;
$$
\begin{equation}
\rho \simeq 0~~:~~~\eta \simeq 0.3 \times 2^{\pm 1}~~~.
\end{equation}

The only evidence for $\eta \ne 0$ comes from CP violation in the kaon system.
We see from Fig.~3 that an acceptable description of this phenomenon can be
obtained with a wide range of CKM parameters. In the next two sections we
mention some tests that could confirm the picture.

\section{Improved tests}\label{sec:imp}

A few ways to improve information about CKM matrix elements are listed below.
Others (e.g., better measurements of $|V_{cb}|$ or $V_{ub}|$) are mentioned
elsewhere in this conference\cite{stonepas} or in more extensive
reviews.\cite{CKM,RVW}

\subsection{Decay constant information}

If $f_B$ were better known, the indeterminacy in the $(\rho,\eta)$ plane
would be reduced considerably.  We show in Fig.~4 the variation in $\chi^2$
for the fit described in the previous section when $f_B$ is taken to have
a fixed value.  An acceptable fit is obtained for a wide range of values,
with $\chi^2 = 0$ for $f_B = 153$ and 187 MeV.

\begin{figure}
\vspace{4in}
\caption{Variation of $\chi^2$ in a fit to CKM parameters as a function
of $f_B$.}
\end{figure}

The reason for the flat behavior of $\chi^2$ with $f_B$ is illustrated in
Fig.~5.  The dashed line, labeled by values of $f_B$, depicts the $(\rho,\eta)$
value for the solution with minimum $\chi^2$ at each $f_B$.  The product
$|1 - \rho - i \eta| f_B$ is constrained to be a constant by $B^0 - \bar B^0$
mixing.  The product $\eta (1 - \rho)$ is constrained to be constant by
the value of $\epsilon$.  The locus of solutions to these two conditions
lies approximately tangent to the circular arc associated with the
constraint on $|V_{ub}/V_{cb}|$ for a wide range of values of $f_B$.

\begin{figure}
\vspace{3in}
\caption{Locus of points in $(\rho,\eta)$ corresponding to minimum $\chi^2$ for
fixed values of $f_B$.  Circular arcs depict central value and $\pm 1 \sigma$
errors for $|V_{ub}/V_{cb}|$.  Solid dots denote points with $\chi^2 = 0$.}
\end{figure}

The uncertainty in $f_B$ thus becomes a major source of uncertainty in
$\rho$, which will not improve much with better information on
$|V_{ub}/V_{cb}|$.  Fortunately, several estimates of $f_B$ are available,
and their reliability should improve.

{\it Lattice gauge theories} have attempted to evaluate decay constants
for $D$ and $B$ mesons.  A representative set\cite{BLS} is
$$
f_B = 187 \pm 10 \pm 34 \pm 15~~{\rm MeV}~~~,
$$
$$
f_{B_s} = 207 \pm 9 \pm 34 \pm 22~~{\rm MeV}~~~,
$$
$$
f_D = 208 \pm 9 \pm 35 \pm 12~~{\rm MeV}~~~,
$$
\begin{equation}
f_{D_s} = 230 \pm 7 \pm 30 \pm 18~~{\rm MeV}~~~,
\end{equation}
where the first errors are statistical, the second are associated with fitting
and lattice constant, and the third arise from scaling from the static $(m_Q =
\infty)$ limit.  The spread between these and some other lattice
estimates\cite{LAT} is larger than the errors quoted above, however.

{\it Direct measurements} are available so far only for the $D_s$ decay
constant.  The WA75 collaboration\cite{WA75} has seen 6 -- 7
$D_s \to \mu \nu$ events and conclude that $f_{D_s} = 232 \pm 69$ MeV.  The
CLEO Collaboration\cite{FDSCLEO} has a much larger statistical sample; the main
errors arise from background subtraction and overall normalization (which
relies on the $D_s \to \phi \pi$ branching ratio).  Using several methods to
estimate this branching ratio, Muheim and Stone\cite{MS} estimate $f_{D_s} =
315 \pm 45$ MeV.  We average this with the WA75 value to obtain $f_{D_s} = 289
\pm 38$ MeV.

{\it Quark models} can provide estimates of decay constants and their ratios.
In a non-relativstic model,\cite{ES} the decay constant $f_M$ of a heavy meson
$M = Q \bar q$ with mass $M_M$ is related to the square of the $Q \bar q$ wave
function at the origin by $f_M^2 = 12 |\Psi(0)|^2 / M_M$.  The ratios of
squares of wave functions can be estimated from strong hyperfine splittings
between vector and pseudoscalar states, $\Delta M_{\rm hfs} \propto
|\Psi(0)|^2/m_Q m_q$.  The equality of the $D_s^* - D_s$ and $D^* - D$
splittings then suggests that
\begin{equation}
f_D/f_{D_s} \simeq (m_d/m_s)^{1/2} \simeq 0.8 \simeq f_B/f_{B_s}~~~,
\end{equation}
where we have assumed that similar dynamics govern
the light quarks bound to charmed and $b$ quarks.  Using our average for
$f_{D_s}$, we find $f_D = (231 \pm 31)$ MeV\null.  This is to be compared
with the Mark III upper limit\cite{MKIII} $f_D < 290$ MeV (90\% c.l.).

An absolute estimate of $|\Psi(0)|^2$ can been obtained using electromagnetic
hyperfine splittings,\cite{AM} which are probed by comparing isospin splittings
in vector and pseudoscalar mesons.  Before corrections of order $(1/m_Q)$,
the values $f_D^{(0)} = (290 \pm 15)$ MeV, $f_B^{(0)} = (177 \pm 9)$
MeV were obtained.  With $f_M = f_M^{(0)} (1 - \Delta/m_M)$ $(M = D,~B)$, we
use our value of $f_D$ to estimate $\Delta/M_D = 0.20 \pm 0.11$, $\Delta/M_B =
0.07 \pm 0.04$, and hence $f_B = f_B^{(0)} (1- \Delta/m_B) = 164 \pm 11$
MeV\null.  Applying a QCD correction\cite{VSPW} of 1.10 to the ratio $f_B/f_D$,
we finally estimate $f_B = (180 \pm 12)$ MeV. [This is the basis of the value
taken in the previous Section, where we inflated the error arbitrarily.] We
also obtain $f_{B_s} = (225 \pm 15)$ MeV from the ratio based on the quark
model.

\subsection{Rates and ratios}

The partial width $\Gamma(B \to \ell \nu)$ is proportional to $f_B^2
|V_{ub}|^2$.  The expected branching ratios are about $(1/2) \times
10^{-4}$ for $\tau \nu$ and $2 \times 10^{-7}$ for $\mu \nu$.
Another interesting ratio\cite{ALI} is $\Gamma(B \to \rho \gamma)/ \Gamma(B
\to K^* \gamma)$, which, aside from small phase space corrections, should just
be $|V_{td}/V_{ts}|^2 \simeq 1/20$.

\subsection{The $K^+ \to \pi^+ \nu \bar \nu$ rate}

The decay $K^+ \to \pi^+ \nu \bar \nu$ is governed by loop diagrams
involving the cooperation of charmed and top quark contributions.  The
branching ratio will provide useful information on $\rho$.\cite{HR}
The ranges of $(\rho,\eta)$ favored in the fit of Fig.~3 imply that for $m_t =
(160,~175,~190)$ GeV/$c^2$ the 90\% c.l. limits of the branching ratio (in
units of $10^{-10}$) are (0.6 to 1.6, 0.7 to 1.9, 0.7 to 2.2).  Thus, the
favored value is slightly above $10^{-10}$, give or take a factor of 2. A low
value within this range signifies $\rho > 0$, while a high value signifies
$\rho < 0$. The present published upper limit\cite{E787} is $B(K^+ \to \pi^+
\nu \bar \nu) < 5 \times 10^{-9}$ (90\% c.l.), with a recent unpublished
improvement\cite{Turc} to $3 \times 10^{-9}$ (90\% c.l.).

\subsection{Other $K_L \to \pi^0 \ell \bar \ell$ decays}

{\it The decay $K_L \to \pi^0 e^+ e^-$} is expected to be dominated by
CP-violating contributions, though the possibility of CP-conserving
contributions through a two-photon intermediate state could be laid to rest by
more detailed studies of the decay $K_L \to \pi^0 \gamma \gamma$.\cite{pigg}
Two types of CP-violating contributions are expected:  ``indirect,'' via the
CP-positive component $K_1$ component of $K_L = K_1 + \epsilon K_2$, and
``direct,'' whose presence would be a detailed verification of the CKM theory
of CP violation.

The indirect contribution is expected to lead to a branching
ratio\cite{RVW,Dib} $B^{\rm in} \simeq 6 \times 10^{-12}$, though one
calculation\cite{Ko} finds it an order of magnitude larger.  The phase of the
indirect contribution is expected to be that of $\epsilon$ (i.e., about
45$^{\circ}$) modulo $\pi$. The direct contribution should be proportional to
$i \eta$, and should be comparable in magnitude to the indirect contribution in
most estimates.  One potential background\cite{HG} is the process $K_L \to
\gamma_1 e^+ e^-$, where the positron or electron radiates a photon
$(\gamma_2)$.  If $m(\gamma_1 \gamma_2)$ is too close to $m_{\pi^0}$, this
process can be confused with the signal.

The present 90\% c.l. upper limit to $B(K_L \to \pi^0 e^+ e^-)$ is
$1.8 \times 10^{-9}$, where results from several experiments have been
combined.\cite{DHE}

{\it The decay $K_L \to \pi^0 \mu^+ \mu^-$} should have less background than
$K_L \to \pi^0 e^+ e^-$ from photons radiated by the charged leptons, and
should have a comparable rate (aside from phase space differences).  The
present 90\% c.l. upper limit\cite{DHM} is $5.1 \times 10^{-9}$.

{\it The decay $K_L \to \pi^0 \nu \bar \nu$} should have a branching ratio
of about $2 \times 10^{-10} \eta^2$ for $m_t = 175$ GeV.  It should have
only a very small indirect contribution, and would be incontrovertible
evidence for the CKM theory.  At present, experimental bounds\cite{pinunu}
are $B(K_L \to \pi^0 \nu \bar \nu) < 5.8 \times 10^{-5}$ (90\% c.l., summed
over neutrino species).

\subsection{The ratio $\epsilon'/\epsilon$ for kaons}

The observation of a nonzero value of $\epsilon'/\epsilon$ (Section 2.2)
has long been viewed as one of the most promising ways to disprove a
``superweak'' theory of this effect.\cite{sw,RVW}

The latest estimates\cite{BEPS} are equivalent (for a top mass of about 170
GeV/$c^2$) to $[\epsilon'/\epsilon]|_{\rm kaons} = (6 \pm 3) \times 10^{-4}
\eta$, with an additional factor of 2 uncertainty associated with hadronic
matrix elements. The Fermilab E731 Collaboration\cite{Gib} measures
$\epsilon'/\epsilon = (7.4 \pm 6) \times 10^{-4}$, leading to no restrictions
on $\eta$ in comparison with the range (0.2 to 0.6) we have already specified.
The CERN NA31 Collaboration\cite{NA31} finds $\epsilon'/\epsilon = (23 \pm
7) \times 10^{-4}$, which is higher than theoretical expectations. Both groups
are preparing new experiments, for which results should be available around
1996.

\section{Detailed $B$ studies}\label{sec:Bdecays}

The properties of $B$ mesons are particularly approriate for testing the
CKM theory of CP violation, since the third family of quarks is essential
to this theory.  Indeed, many aspects of $B$ physics provide relevant
information, including $B_s - \bar B_s$ mixing, CP violation in neutral and
charged $B$ decays, and the study of relations among decay rates.

\subsection{$B_s - \bar B_s$ mixing}

The same type of box diagrams which lead to $K^0 - \bar K^0$ and $B^0 - \bar
B^0$ mixing also mix strange $B$ mesons with their antiparticles. One expects
$(\Delta m)|_{B_s}/(\Delta m)|_{B_d} = (f_{B_s}/f_{B_d})^2 (B_{B_s}/B_{B_d})
|V_{ts}/V_{td}|^2$, which should be a very large number (of order 20 or more).
Thus, strange $B$'s should undergo many particle-antiparticle oscillations
before decaying.

The main uncertainty in an estimate of $x_s \equiv (\Delta m/
\Gamma)_{B_s}$ is associated with $f_{B_s}$.  The CKM elements $V_{ts} \simeq
-0.04$ and $V_{tb} \simeq 1$ which govern the dominant (top quark) contribution
to the mixing are known reasonably well. We show in Table 1 the dependence of
$x_s$ on $f_{B_s}$ and $m_t$. To measure $x_s$, one must study the
time-dependence of decays to specific final states and their charge-conjugates
with resolution equal to a small fraction of the $B_s$ lifetime (about 1.5 ps).

\begin{table}
\begin{center}
\caption{Dependence of mixing parameter $x_s$ on top quark mass and
$B_s$ decay constant.}
\medskip
\begin{tabular}{c c c c} \hline
\null \qquad $m_t$ (GeV/$c^2$)&  157  &  174  &  191  \\ \hline
$f_{B_s}$ (MeV)               &       &       &       \\
150                           &  7.6  &  8.9  & 10.2  \\
200                           & 13.5  & 15.8  & 18.2  \\
250                           & 21.1  & 24.7  & 28.4  \\ \hline
\end{tabular}
\end{center}
\end{table}

\subsection{CP violation in $B$ decays}

Soon after the discovery of the $\Upsilon$ states it was realized that
CP-violating phenomena in decays of $B$ mesons were expected to be observable
and informative.\cite{EGNR,BCP}

{\it Decays of neutral $B$ mesons to CP eigenstates $f$} can exhibit rate
asymmetries (or time-dependent asymmetries) as a result of the interference
of the direct process $B^0 \to f$ and the two-step process $B^0 \to \bar B^0
\to f$ involving mixing.

{\it Decays to CP non-eigenstates} can exhibit rate asymmetries only if
there are two different weak decay amplitudes {\it and} two different
strong phase shifts associated with them.  Comparing
\begin{equation}
\Gamma \equiv \Gamma(B^0 \to f)
= |a_1 e^{i(\phi_1 + \delta_1)} +  a_2 e^{i(\phi_2 + \delta_2)}|^2
\end{equation}
with
\begin{equation}
\bar \Gamma \equiv \Gamma(\bar B^0 \to \bar f)
= |a_1 e^{i(-\phi_1 + \delta_1)} +  a_2 e^{i(-\phi_2 + \delta_2)}|^2
\end{equation}
we notice that only the weak phases $\phi_i$ change sign under CP inversion,
not the strong phases $\delta_i$.  Defining $\phi \equiv \phi_1 - \phi_2$,
$\delta \equiv \delta_1 - \delta_2$, we find
\begin{equation}
\Gamma - \bar \Gamma \propto 2 a_1 a_2 \sin \phi \sin \delta~~~,
\end{equation}
so both $\phi$ and $\delta$ must be nonzero in order to see a rate difference.

\subsection{Decays involving neutral $B$'s}

Asymmetries decays of neutral $B$ mesons to CP eigenstates can provide direct
information about the angles in the unitarity triangle of Fig.~1.
We may define a time-integrated rate asymmetry $A(f)$ as
\begin{equation}
A(f) \equiv \frac{\Gamma(B^0_{t=0} \to f) - \Gamma(\bar B^0_{t=0} \to f)}
{\Gamma(B^0_{t=0} \to f) + \Gamma(\bar B^0_{t=0} \to f)}~~~.
\end{equation}
The angles $\beta$ and $\alpha$ in Fig.~1 are related to the asymmetries
in decays to $J/\psi K_S$ and $\pi^+ \pi^-$ final states:
\begin{equation}
A(J/\psi K_S) = - \frac{x_d}{1 + x_d^2} \sin 2 \beta~~~,
\end{equation}
\begin{equation}
A(\pi^+ \pi^-) = - \frac{x_d}{1 + x_d^2} \sin 2 \alpha~~~,
\end{equation}
where $x_d \equiv (\Delta m/\Gamma)|_{B^0}$, and we have neglected lifetime
differences between eigenstates.  The contours of Fig.~3 imply that
the asymmetry $A(J/\psi K_S)$ should be between $-0.1$ and
its most negative value of $x_d^2/(1 + x_d^2) = -0.47$.  If it lay outside
these bounds, it could immediately disprove the CKM explanation of CP violation
in the neutral kaon system.  The asymmetry $A(\pi^+ \pi^-)$ can take on any
value between $-0.47$ and $+0.47$.  It would thus provide very helpful
information on CKM parameters.

\subsection{Tagging of neutral $B$ decays}

In the study of decays of neutral $B$ mesons to CP eigenstates, one must know
whether a $B^0$ or $\bar B^0$ was produced at $t = 0$. One suggestion\cite{GNR}
is to correlate the neutral $B$ meson with a charged pion, as in the decay of a
higher-lying resonance. This tagging method\cite{SN} has been used to tag
neutral charmed mesons, where the decays $D^{*+} \to D^0 \pi^+$ and $D^{*-} \to
\bar D^0 \pi^-$ {\it are} kinematically allowed, in order to look for mixing or
to compare decays involving the highly suppressed subprocess $c \to d u \bar s$
with those involving the favored subprocess $c \to s u \bar d$.

The usefulness of the method for $B$ mesons can be checked by learning the
degree to which a correlation exists between a charged pion and a $B^0$ decay
to states of {\it identified flavor} such as $J/\psi K^{*0}$ (with $K^{*0} \to
K^+ \pi^-$) or $D^- \pi^+$.  Under normal circumstances the $B$ mesons are
produced in an isospin-independent manner and one can study these correlations
using charged $B$ mesons as well.

The correlation is easily visualized with the help of quark diagrams. By
convention (the same as for kaons), a neutral $B$ meson containing an initially
produced $\bar b$ is a $B^0$.  It also contains a $d$ quark.  The next charged
pion down the fragmentation chain must contain a $\bar d$, and hence must be a
$\pi^+$.  Similarly, a $\bar B^0$ will be correlated with a $\pi^-$.

The same conclusion can be drawn by noting that a $B^0$ can resonate with a
positive pion to form an excited $B^+$, which we shall call $B^{**+}$ (to
distinguish it from the $B^*$, lying less than 50 MeV/$c^2$ above the $B$).
Similarly, a $\bar B^0$ can resonate with a negative pion to form a $B^{**-}$.
The combinations $B^0 \pi^-$ and $\bar B^0 \pi^+$ are {\it exotic},~i.e., they
cannot be formed as quark-antiquark states.  No evidence for exotic resonances
exists.

The lightest states which can decay to $B \pi$ and/or $B^* \pi$ are P-wave
resonances of a $b$ quark and an $\bar u$ or $\bar d$. The expectations for
masses of these states\cite{GNR,EHQ}, based on extrapolation from the known
$D^{**}$ resonances, are summarized in Table 2.

\begin{table}
\begin{center}
\caption{P-wave resonances of a $b$ quark and a light ($\bar u$
or $\bar d$) antiquark}
\medskip
\begin{tabular}{c c c} \hline
$J^P$  &    Mass     &  Allowed final \\
       & (GeV/$c^2$) &     state(s)   \\ \hline
$2^+$  & $\sim 5.77$ &  $B \pi,~B^* \pi$ \\
$1^+$  & $\sim 5.77$ &     $B^* \pi$  \\
$1^+$  & $ < 5.77$   &     $B^* \pi$  \\
$0^+$  & $ < 5.77$   &      $B \pi$   \\ \hline
\end{tabular}
\end{center}
\end{table}

The known $D^{**}$ resonances are a $2^+$ state around 2460 MeV/$c^2$, decaying
to $D \pi$ and $D^* \pi$, and a $1^+$ state around 2420 MeV/$c^2$, decaying to
$D^* \pi$.  These states are relatively narrow, probably because they decay
via a D-wave.  In addition, there are expected to be much broader (and probably
lower) $D^{**}$ resonances:  a $1^+$ state decaying to $D^* \pi$ and a $0^+$
state decaying to $D \pi$, both via S-waves.

\subsection{Decays of $B$'s to pairs of pseudoscalars}

I would like to report on some new work in collaboration with M. Gronau, O.
Hern\'andez, and D. London.\cite{BPP}  The decays $B \to (\pi \pi, \pi K, K
\bar K)$ are a rich source of information on both weak (CKM) and strong phases,
if we are willing to use flavor SU(3) symmetry.

The decays $B \to \pi \pi$ are governed by transitions $b \to d q \bar q~(q =
u,~d, \ldots)$ with $\Delta I = 1/2$ and $\Delta I = 3/2$, leading respectively
to final states with $I = 0$ and $I = 2$.  Since there is a single amplitude
for each final isospin but three different charge states in the decays, the
amplitudes obey a triangle relation:  $A(\pi^+ \pi^-) - \sqrt{2} A(\pi^0 \pi^0)
= \sqrt{2}  A(\pi^+ \pi^0)$.  The triangle may be compared with that for the
charge-conjugate processes and combined with information on time-dependent $B
\to \pi^+ \pi^-$ decays to obtain information on weak phases.\cite{pipi}

The decays $B \to \pi K$ are governed by transitions $b \to s q \bar q~(q =
u,~d, \ldots)$ with $\Delta I = 0$ and $\Delta I = 1$.  The $I = 1/2$ final
state can be reached by both $\Delta I = 0$ and $\Delta I = 1$ transitions,
while only $\Delta I = 1$ contributes to the $I = 3/2$ final state.
Consequently, there are three independent amplitudes for four decays, and
one quadrangle relation $A(\pi^+ K^0) + \sqrt{2}A(\pi^0 K^+) = A(\pi^- K^+)
+ \sqrt{2} A(\pi^0 K^0)$.  As in the $\pi \pi$ case, this relation may be
compared with the charge-conjugate one and the time-dependence of decays to CP
eigenstates (in this case $\pi^0 K_S$) studied to obtain CKM phase
information.\cite{pik}

We have re-examined\cite{BPP} SU(3) analyses\cite{OldSU} of the decays $B \to P
P$ ($P$ = light pseudoscalar). They imply a number of useful relations among
$\pi \pi,~\pi K$, and $K \bar K$ decays, among which is one relating $B^+$
amplitudes alone:
\begin{equation}
A(\pi^+ K^0) + \sqrt{2} A(\pi^0 K^+) = \tilde{r}_u \sqrt{2} A(\pi^+ \pi^0)~~~.
\end{equation}
Here $\tilde{r}_u \equiv (f_K/f_\pi)|V_{us}/V_{ud}|$.  This expression relates
one side of the $\pi \pi$ amplitude triangle to one of the diagonals of the
$\pi K$ amplitude quadrangle, and thus reduces the quadrangle effectively to
two triangles, simplifying previous analyses.\cite{pik}  Moreover, since
one expects the $\pi^+ K^0$ amplitude to be dominated by a penguin diagram
(with expected weak phase $\pi$) and the $\pi^+ \pi^0$ amplitude to have the
phase $\gamma = {\rm Arg}~V_{ub}^*$, the comparison of this last relation
and the corresponding one for charge-conjugate decays can provide information
on the weak phase $\gamma$.

In Fig.~6 we compare the amplitude triangles for $B^+$ and $B^-$ decays.  The
angle between the sides corresponding to $B^+ \to \pi^+ \pi^0$ and $B^- \to
\pi^- \pi^0$ is $2 \gamma$. Two possible orientations for the triangles are
shown.
\begin{figure}
\vspace{7in}
\caption{Amplitude triangles illustrating the extraction of the weak
phase $\gamma$ from charged $B$ decays to $\pi \pi$ and $\pi K$.}
\end{figure}

The amplitude $A(\pi^0 K^+)$ which is the sum of ``tree'' and penguin graph
contributions can be expressed\cite{LWC} as
\begin{equation}
\sqrt{2}A(\pi^0 K^+) = A e^{i \gamma} e^{i \delta_3} + B e^{i \delta_P}
= e^{i \delta_P} (A e^{i \gamma} e^{i \delta} + B)~~,
\end{equation}
where $\delta_3$ is the strong $I = 3/2$ phase, $\delta_P$ is the strong
penguin phase, and $\delta \equiv \delta_3 - \delta_P$.  In the corresponding
expression for $A(\pi^0 K^-)$, only the sign of $\gamma$ is changed.  The
$\pi^0 K^{\pm}$ decay rates are then
\begin{equation}
\Gamma(\pi^0 K^{\pm}) = (1/2)[A^2 + B^2 + 2 A B \cos(\delta \pm \gamma)]~~~.
\end{equation}
The sum and difference of these rates are
\begin{equation}
S \equiv \Gamma(\pi^0 K^+) + \Gamma(\pi^0 K^-) = A^2 + B^2 + 2AB \cos \delta
\cos \gamma~~~,
\end{equation}
\begin{equation}
D \equiv \Gamma(\pi^0 K^+) - \Gamma(\pi^0 K^-) = 2 A B \sin \delta \sin
\gamma~~~.
\end{equation}
The CP-violating rate difference $D$ is probably very small as a result of
the likely smallness of the phase difference $\delta$.

Let us take as an example $|A/B| = 1/3,~\delta = 0$, and $\cos \gamma \simeq
0$. Expressing $\cos \gamma = (S - A^2 - B^2)/2AB$, we find that in order to
measure $\gamma$ to $10^{\circ}$ one needs a sample consisting of about 100
events in the channel $\pi^0 K^{\pm}$ corresponding to $S$.

Further relations can be obtained\cite{BPP} by comparing the amplitude
triangles involving both charged and neutral $B$ decays to $\pi K$. By looking
at the amplitude triangles for these decays and their charge conjugates, one
can sort out a number of weak {\it and} strong phases.

Some combination of the decays $B^0 \to \pi^+ \pi^-$ and $B^0 \to \pi^- K^+$
has already been observed.\cite{Battle}  The sum of these two modes is nonzero
at better than the $4 \sigma$ level, corresponding to a combined branching
ratio of $2 \times 10^{-5}$.  The most likely solution is that there are about
7 events in each channel, each with a branching ratio of about $10^{-5}$.
Since the $B^0 \to \pi^+ \pi^-$ decay is expected to be dominated by a ``tree''
subprocess $b \to u \bar u d$ while $B^0 \to \pi^- K^+$ should be dominated by
a penguin amplitude $b \to s +$ gluon, there is hope that these amplitudes
should be of comparable size.  Taking account of the likely size of the factor
$\tilde{r}_u$ then led us to the estimate $|A/B| = 1/3$ mentioned above.

Other 90\% c.l. upper limits\cite{Wuert} are $B(B^0 \to \pi^0 \pi^0) < 2.7
\times 10^{-5}$, $B(B^+ \to \pi^+ \pi^0) < 4.4 \times 10^{-5}$, and $B(B^+ \to
\pi^0 K^+) < 2.6 \times 10^{-5}$.  It will be necessary to improve the data
sample by about a factor of 100 before the tests mentioned above can be
contemplated, but this is within the possibility of planned facilities.

\section{Alternative sources of CP violation}\label{sec:alt}

\subsection{The superweak model}

It is possible to explain the nonzero value of $\epsilon$ in the neutral kaon
system by means of an {\it ad hoc} $\Delta S = 2$ interaction leading directly
to CP-violating $K^0 - \bar K^0$ mixing.\cite{sw}  The phase of $\epsilon$ will
then automatically be the superweak phase mentioned in Section 2, and one will
see no difference between $\eta_{+-}$ and $\eta_{00}$. The only evidence
against this possibility so far is the $3 \sigma$ observation of nonzero
$\epsilon'/\epsilon$ by the CERN NA31 experiment,\cite{NA31} a result not
confirmed by Fermilab E731.\cite{Gib}

A superweak interaction (of considerably greater strength) could in principle
lead to observable CP-violating $B^0 - \bar B^0$ mixing.  If this were so,
one would expect\cite{Win} $A(\pi^+ \pi^-) = - A(J/\psi K_S)$ as a result of
the opposite CP eigenvalues of the two final states.  In order for this
relation to hold in the standard model, one would need $\eta = (1-\rho)
[\rho/(2-\rho)]^{1/2}$.  Taking account of possible errors in checking that
the asymmetries are actually equal and opposite, one concludes that a portion
of the allowed region of parameters shown in Fig.~3 could not be distinguished
from a superweak theory.  The ratio $A(\pi^+ \pi^-)/A(J/\psi K_S)$ is
informative in a more general context:  for example, if it exceeds
1, then $\rho$ must be negative.\cite{PHJR}

If $\epsilon$ arises entirely from a superweak interaction, there is no need
for CKM phases, and one will see no ``direct'' effects in kaon or $B$ decays.
There will also be no neutron or electron electric dipole moments, though such
effects also will be well below experimental capabilities in the standard CKM
picture.

\subsection{Right-handed $W$'s}

If there are new $W$ bosons (``right-handed $W$'s, or $W_R$) coupling to
right-handed fermions, one can obtain CP-violating interactions (for example,
via box diagrams involving $W_R$ and the usual left-handed $W_L$).  The
right-handed $W$ mass scale must be tens of TeV or less in order for a large
enough contribution to $\epsilon$ to be generated.  In contrast to the
situation described in Section 3.2, one can generate CP violation using only
two quark families, since redefinitions of quark phases are constrained by the
right-handed couplings.

An amusing feature of right-handed $W$ couplings is that their participation
(or even dominance) in $b$ quark decays is surprisingly hard to rule
out.\cite{GW}  Some suggestions have been made to test the usual picture of
left-handed $b \to c$ decays using the polarization of $\Lambda_b$ baryons
produced in $b$ quark fragmentation.\cite{ARWW}

\subsection{Multi-Higgs models}

If there is more than one Higgs doublet, complex vacuum expectation values of
Higgs fields can lead to CP-violating effects.\cite{LW}  It appears that in
order to explain $\epsilon \ne 0$ in neutral kaon decays by this mechanism, one
expects too large a neutron electric dipole moment.\cite{Sanda}  The
possibility of such effects remains open, however, and the best test for them
remains the study of dipole moments.  Current models\cite{Hay,YLWu} tend to be
constrained by the present limits\cite{limits} of
\begin{equation}
|d_n| < 1.1 \times 10^{-25}~e{\cdot \rm cm}~(95\%~{\rm c.l.})~~,~~~
|d_e| < 2 \times 10^{-26}~e{\cdot \rm cm}~(95\%~{\rm c.l.})~~.
\end{equation}
Other CP-violating effects in multi-Higgs-boson models include transverse
lepton polarization in $K_{\mu 3}$ decays\cite{FKG} and various asymmetries
in charm decays,\cite{YLWu} which we now discuss briefly.

\section{Charm decays}\label{sec:cha}

The standard model predicts very small CP-violating effects in charmed particle
decays.  $D^0 - \bar D^0$ mixing is expected to be small and uncertain,
dominated by long-distance effects.\cite{Dmix}  The short-distance contribution
to CP-violating mixing should be of order $(m_b/m_t)^2$ times that in neutral
kaons, while the lifetime of a neutral $D$ meson is about 0.4 ps in comparison
with 52 ns for a neutral kaon.  The tree-level decays $c \to s u \bar d$, $c
\to d u \bar d$, $c \to d u \bar s$ have zero or negligible weak phases in the
standard convention.

For precisely these reasons, CP-violating charmed particle decays are an
excellent place to look for new physics.\cite{YLWu,BigC}  New effects tend to
be accompanied with flavor-changing neutral currents, which may or may not be
an advantage in specific cases.  Experiments are easy to perform and
undersubscribed in comparison with the many proposed studies of $B$ physics.
Information on rate asymmetries $A(f) \equiv [\Gamma(i \to f) - \Gamma(\bar{i}
\to \bar f)]/[\Gamma(i \to f) + \Gamma(\bar{i} \to \bar f)]$ is just now
beginning to appear, as illustrated in Table 3.\cite{Asymms}

\begin{table}
\begin{center}
\caption{Rate asymmetries in charmed meson decays.}
\bigskip
\begin{tabular}{c c c} \hline
Charmed &      Final       &          Asymmetry \\
meson   &      state       &                    \\ \hline
$D^+$   & $K^-K^+\pi^+$    & $-0.031 \pm 0.068$ \\
$D^+$   & $\bar K^{*0}K^+$ & $-0.12 \pm 0.13$   \\
$D^+$   & $\phi \pi^+$     & $0.066 \pm 0.086$  \\
$D^0$   &    $K^+ K^-$     & $0.024 \pm 0.084$  \\ \hline
\end{tabular}
\end{center}
\end{table}
\bigskip

\section{Baryogenesis}\label{sec:bau}

The ratio of baryons to photons in our Universe is a few parts in $10^9$. If
baryons and antibaryons had been produced in equal numbers, mutual
annihilations should have reduced this quantity to a much smaller
number,\cite{JP} of order a part in $10^{18}$. In 1967 Sakharov\cite{Sakh}
proposed three ingredients of any theory which sought to explain the
preponderance of baryons over antibaryons in our Universe:  (1) violation of C
and CP; (2) violation of baryon number, and (3) a period in which the Universe
was out of thermal equilibrium.  Thus our very existence may owe itself to CP
violation.  However, no consensus exists on a specific implementation of
Sakharov's suggestion.

A toy model illustrating Sakharov's idea can be constructed within an SU(5)
grand unified theory.  The gauge group SU(5) contains ``$X$'' bosons which can
decay both to $uu$ and to $e^+ \bar d$.  By CPT, the total decay rates of $X$
and $\bar X$ must be equal, but CP-violating rate differences $\Gamma(X \to uu)
\ne \Gamma(\bar X \to \bar u \bar u)$ and $\Gamma(X \to e^+ \bar d) \ne
\Gamma(\bar X \to e^- d)$ are permitted.  This example conserves $B - L$, where
$B$ is baryon number (1/3 for quarks) and $L$ is lepton number (1 for
electrons).

It was pointed out by 't Hooft\cite{tH} that the electroweak theory contains an
anomaly as a result of nonperturbative effects which conserves $B - L$ but
violates $B + L$.  If a theory leads to $B - L = 0$ but $B + L \ne 0$ at some
primordial temperature $T$, the anomaly can wipe out any $B+L$ as $T$ sinks
below the electroweak scale.\cite{KRS}  Thus, the toy model mentioned above and
many others are unsuitable in practice.  Proposed solutions include (1) the
generation of baryon number directly at the electroweak scale rather than at a
higher temperature,\cite{FS} and (2) the generation of nonzero $B - L$ at a
high temperature, e.g., through the generation of nonzero lepton number $L$
which is then reprocessed into nonzero baryon number by the `t Hooft anomaly
mechanism.\cite{Yana} The first scenario, based on standard model CP-violating
interactions (as manifested in the CKM matrix), is widely regarded as
inadequate to generate the observed baryon asymmetry at the electroweak
scale.\cite{HS}  We illustrate in Fig.~7 some aspects of the second scenario.
The existence of a baryon asymmetry, when combined with information on
neutrinos, could provide a window to a new scale of particle physics.

\begin{figure}
\vspace{5in}
\caption{Mass scales associated with one scenario for baryogenesis.}
\end{figure}

If neutrinos have masses at all, they are much lighter than their charged
counterparts or the corresponding leptons.  One possibility for the suppression
of neutrino masses\cite{seesaw} is the so-called ``seesaw'' mechanism, by which
light neutrinos acquire Majorana masses of order $m_M = m_D^2/M_M$, where $m_D$
is a typical Dirac mass and $M_M$ is a large Majorana mass acquired by
right-handed neutrinos.  Such Majorana masses change lepton number by two units
and therefore are ideal for generating a lepton asymmetry if Sakharov's other
two conditions are met.

The question of baryogenesis is thus shifted onto the leptons:  Do neutrinos
indeed have masses?  If so, what is their ``CKM matrix''?  Do the properties of
heavy Majorana right-handed neutrinos allow any new and interesting natural
mechanisms for violating CP at the same scale where lepton number is violated?
Majorana masses for right-handed neutrinos naturally violate left-right
symmetry and could be closely connected with the violation of $P$ and $C$ in
the weak interactions.\cite{BKCP}

An open question in this scenario, besides the precise form of CP violation at
the lepton-number-violating scale, is how this CP violation gets communicated
to the lower mass scale at which we see CKM phases.  Presumably this occurs
through higher-dimension operators which imitate the effect of Higgs boson
couplings to quarks and leptons.

\section{The strong CP problem}\label{sec:scp}

We did not have time to present this material orally, and so will be very brief
in the written version.  There are fine reviews elsewhere.\cite{RP,MD}

As a result of nonperturbative effects, the QCD Lagrangian acquires an added
CP-violating term $g_s^2 \bar \theta F_{\mu \nu}^a \tilde{F}^{\mu \nu a}/32
\pi^2$, where $\bar \theta = \theta + {\rm Arg~det}~M$, $\theta$ is a term
describing properties of the QCD vacuum, and $M$ is the quark mass matrix. The
limit on the observed neutron electric dipole moment, together with the
estimate\cite{RP} $d_n \simeq 10^{-16} \bar \theta~e$ cm, implies that
$\theta \le 10^{-9}$, which looks very much like zero.  How can one undertand
this?  Several proposals exist.

\subsection{Vanishing $m_u$}

If one quark mass vanishes (the most likely candidate being $m_u$), one can
rotate away any effects of $\bar \theta$.\cite{RP}  However, it is
generally though not universally felt that this bends the constraints of
chiral symmetry beyond plausible limits.\cite{Leut}  My own guess is that
light-quark masses are in the ratios $u:d:s = 3:5:100$.

\subsection{Axion}

One can introduce a continuous U(1) global symmetry\cite{PQ} such that
$\bar \theta$ becomes a dynamical variable which relaxes to zero as a result of
new interactions.  The spontaneous breaking of this symmetry then leads to
a pseudo-Nambu-Goldstone boson, the {\it axion},\cite{AX} for which searches
may be performed in many ways.  My favorite is via the Primakoff
effect,\cite{PS} in which axions in the halo of our galaxy interact with a
static man-made strong magnetic field to produce photons with frequency
equal to the axion mass (up to Doppler shifts).  These photons can be detected
in resonant cavities.  Present searches would have to be improved by about
a factor of 100 to detect axions playing a significant role in the mass of the
galaxy.\cite{MD}

\subsection{Boundary conditions}

It has been proposed\cite{RGS} that one consider not the $\theta$-vacuum, but
an incoherent mixture consisting of half $\theta$ and half $-\theta$, in the
manner of a sum over initial spins of an unpolarized particle.  The
experimental consequences of this proposal are still being worked out.

\section{Summary}\label{sec:sum}

The observed CP violation in the neutral kaon system has been successfully
parametrized in terms of the Cabibbo-Kobayashi-Maskawa (CKM) matrix. The
problem has been shifted to one of understanding the magnitudes and phases of
CKM elements.  Even before this more ambitious question is addressed, however,
one seeks independent tests of the CKM picture of CP violation.  Rare $K$
decays and $B$ decays will provide many such tests.

Alternative (non-CKM) theories of CP violation are much more encouraging for
some CP-violating quantities like the neutron electric dipole moment or effects
in charmed particle decays. However, most of these alternative theories do not
predict observable direct CP-violating effects in $K$ or $B$ decays.

No real understanding exists yet of baryogenesis or of the strong CP problem.
Fortunately, there exist many possibilities for experiments bearing on these
questions, including searches for neutrino masses and for axions.

The CKM picture suggests that we may understand CP violation better when the
pattern of fermion masses itself is understood.  As an example, why is the
top quark heavier than all the other quarks and leptons, or why are the others
so much lighter than the top?  The observation of the top quark\cite{CDFtop,HF}
has finally thrust this problem in our faces, and perhaps we will figure out
the answer some day.

\section{Acknowledgments}

I would like to thank several people for enjoyable collaborations on aspects of
the work presented here:  Geoff Harris on the work which permitted several of
the figures to be drawn, and Alex Nippe, Michael Gronau, David London, and
Oscar Hern\'andez on the topics mentioned in Section 5.  In addition I am
grateful to Isi Dunietz, Harry Lipkin, Bob Sachs, Sheldon Stone, Bruce
Winstein, and Lincoln Wolfenstein for fruitful discussions. This work was
supported in part by the United States Department of Energy under Grant No. DE
FG02 90ER40560.

\section{References}

%
\def \ap#1#2#3{{\it Ann. Phys. (N.Y.)} {\bf#1} (#3) #2}
\def \apny#1#2#3{{\it Ann.~Phys.~(N.Y.)} {\bf#1} (#3) #2}
\def \app#1#2#3{{\it Acta Physica Polonica} {\bf#1} (#3) #2}
\def \arnps#1#2#3{{\it Ann. Rev. Nucl. Part. Sci.} {\bf#1} (#3) #2}
\def \arns#1#2#3{{\it Ann. Rev. Nucl. Sci.} {\bf#1} (#3) #2}
\def \ba88{{\it Particles and Fields 3} (Proceedings of the 1988 Banff Summer
Institute on Particles and Fields), edited by A. N. Kamal and F. C. Khanna
(World Scientific, Singapore, 1989)}
\def \baphs#1#2#3{{\it Bull. Am. Phys. Soc.} {\bf#1} (#3) #2}
\def \be87{{\it Proceedings of the Workshop on High Sensitivity Beauty
Physics at Fermilab,} Fermilab, Nov. 11-14, 1987, edited by A. J. Slaughter,
N. Lockyer, and M. Schmidt (Fermilab, Batavia, IL, 1988)}
\def \cn{Collaboration}
\def \cp89{{\it CP Violation,} edited by C. Jarlskog (World Scientific,
Singapore, 1989)}
\def \dpf91{{\it The Vancouver Meeting - Particles and Fields '91}
(Division of Particles and Fields Meeting, American Physical Society,
Vancouver, Canada, Aug.~18-22, 1991), ed. by D. Axen, D. Bryman, and M. Comyn
(World Scientific, Singapore, 1992)}
\def \dpff{{\it The Fermilab Meeting - DPF 92} (Division of Particles and
Fields
Meeting, American Physical Society, Fermilab, 10 -- 14 November, 1992), ed. by
C. H. Albright \ite~(World Scientific, Singapore, 1993)}
\def \efi{Enrico Fermi Institute Report No.~}
\def \hb87{{\it Proceeding of the 1987 International Symposium on Lepton and
Photon Interactions at High Energies,} Hamburg, 1987, ed. by W. Bartel
and R. R\"uckl (Nucl.~Phys.~B, Proc. Suppl., vol. 3) (North-Holland,
Amsterdam, 1988)}
\def \ib{{\it ibid.}~}
\def \ibj#1#2#3{{\it ibid.} {\bf#1} (#3) #2}
\def \ijmpa#1#2#3{{\it Int.~J. Mod.~Phys.}~A {\bf#1} (#3) #2}
\def \ite{{\it et al.}}
\def \jpg#1#2#3{{\it J. Phys.} G {\bf#1} (#3) #2}
\def \kdvs#1#2#3{{\it Kong.~Danske Vid.~Selsk., Matt-fys.~Medd.} {\bf #1}
(#3) No #2}
\def \ky85{{\it Proceedings of the International Symposium on Lepton and
Photon Interactions at High Energy,} Kyoto, Aug.~19-24, 1985, edited by M.
Konuma and K. Takahashi (Kyoto Univ., Kyoto, 1985)}
\def \lat90{{\it Results and Perspectives in Particle Physics} (Proceedings of
Les Rencontres de Physique de la Vallee d'Aoste [4th], La Thuile, Italy, Mar.
18-24, 1990), edited by M. Greco (Editions Fronti\`eres, Gif-Sur-Yvette,
France,
1991)}
\def \lg91{International Symposium on Lepton and Photon Interactions, Geneva,
Switzerland, July, 1991}
\def \lkl87{{\it Selected Topics in Electroweak Interactions} (Proceedings of
the Second Lake Louise Institute on New Frontiers in Particle Physics, 15 --
21 February, 1987), edited by J. M. Cameron \ite~(World Scientific, Singapore,
1987)}
\def \mpla #1#2#3{{\it Mod. Phys. Lett.} A {\bf#1} (#3) #2}
\def \nc#1#2#3{{\it Nuovo Cim.} {\bf#1} (#3) #2}
\def \np#1#2#3{{\it Nucl. Phys.} {\bf#1} (#3) #2}
\def \oxf65{{\it Proceedings of the Oxford International Conference on
Elementary Particles} 19/25 Sept.~1965, ed.~by T. R. Walsh (Chilton, Rutherford
High Energy Laboratory, 1966)}
\def \pisma#1#2#3#4{{\it Pis'ma Zh. Eksp. Teor. Fiz.} {\bf#1} (#3) #2 [{\it
JETP Lett.} {\bf#1} (#3) #4]}
\def \pl#1#2#3{{\it Phys. Lett.} {\bf#1} (#3) #2}
\def \plb#1#2#3{{\it Phys. Lett.} B {\bf#1} (#3) #2}
\def \ppnp#1#2#3{{\it Prog. Part. Nucl. Phys.} {\bf#1} (#3) #2}
\def \pr#1#2#3{{\it Phys. Rev.} {\bf#1} (#3) #2}
\def \prd#1#2#3{{\it Phys. Rev.} D {\bf#1} (#3) #2}
\def \prl#1#2#3{{\it Phys. Rev. Lett.} {\bf#1} (#3) #2}
\def \prp#1#2#3{{\it Phys. Rep.} {\bf#1} (#3) #2}
\def \ptp#1#2#3{{\it Prog. Theor. Phys.} {\bf#1} (#3) #2}
\def \rmp#1#2#3{{\it Rev. Mod. Phys.} {\bf#1} (#3) #2}
\def \rp#1{~~~~~\ldots\ldots{\rm rp~}{#1}~~~~~}
\def \si90{25th International Conference on High Energy Physics, Singapore,
Aug. 2-8, 1990, Proceedings edited by K. K. Phua and Y. Yamaguchi (World
Scientific, Teaneck, N. J., 1991)}
\def \slac75{{\it Proceedings of the 1975 International Symposium on
Lepton and Photon Interactions at High Energies,} Stanford University, Aug.
21-27, 1975, edited by W. T. Kirk (SLAC, Stanford, CA, 1975)}
\def \slc87{{\it Proceedings of the Salt Lake City Meeting} (Division of
Particles and Fields, American Physical Society, Salt Lake City, Utah, 1987),
ed. by C. DeTar and J. S. Ball (World Scientific, Singapore, 1987)}
\def \smass82{{\it Proceedings of the 1982 DPF Summer Study on Elementary
Particle Physics and Future Facilities}, Snowmass, Colorado, edited by R.
Donaldson, R. Gustafson, and F. Paige (World Scientific, Singapore, 1982)}
\def \smass90{{\it Research Directions for the Decade} (Proceedings of the
1990 DPF Snowmass Workshop), edited by E. L. Berger (World Scientific,
Singapore, 1991)}
\def \smassb{{\it Proceedings of the Workshop on $B$ Physics at Hadron
Accelerators}, Snowmass, Colorado, 21 June -- 2 July 1994, ed.~by P. McBride
and C. S. Mishra, Fermilab report FERMILAB-CONF-93/267 (Fermilab, Batavia, IL,
1993)}
\def \stone{{\it B Decays}, edited by S. Stone (World Scientific, Singapore,
1994)}
\def \tasi90{{\it Testing the Standard Model} (Proceedings of the 1990
Theoretical Advanced Study Institute in Elementary Particle Physics),
edited by M. Cveti\v{c} and P. Langacker (World Scientific, Singapore, 1991)}
\def \yaf#1#2#3#4{{\it Yad. Fiz.} {\bf#1} (#3) #2 [Sov. J. Nucl. Phys. {\bf #1}
 (#3) #4]}
\def \zhetf#1#2#3#4#5#6{{\it Zh. Eksp. Teor. Fiz.} {\bf #1} (#3) #2 [Sov.
Phys. - JETP {\bf #4} (#6) #5]}
\def \zhetfl#1#2#3#4{{\it Pis'ma Zh. Eksp. Teor. Fiz.} {\bf #1} (#3) #2 [JETP
Letters {\bf #1} (#3) #4]}
\def \zp#1#2#3{{\it Zeit. Phys.} {\bf#1} (#3) #2}
\def \zpc#1#2#3{{\it Zeit. Phys.} C {\bf#1} (#3) #2}


\end{document}